\documentstyle[aps,prl,twocolumn,epsfig,floats]{revtex}
\input epsf

\begin{document}

\wideabs{
\draft
\title{Finite Temperature Collapse of a Bose Gas with Attractive
Interactions}
\author{
Erich J.~Mueller\cite{REF:EJM}
and Gordon Baym\cite{REF:GB}}
\address{
Department of Physics,
University of Illinois at Urbana-Champaign,
1110 West Green Street,
Urbana, IL 61801}
\date{October 26, 1999}
\maketitle
\begin{abstract}
We study the mechanical stability of the weakly interacting Bose gas with
attractive interactions, and construct a unified picture of the collapse valid
from the low temperature condensed region to the high temperature classical
regime.  As we show, the non-condensed particles play a crucial role in
determining the region of stability.  We extend our results to describe
domain formation in spinor condensates.
\end{abstract}
}
%
%
%

\narrowtext

    The trapped Bose gas with attractive interactions is a novel physical
system.  At high densities these clouds are unstable against collapse; 
however at low densities they can be stabilized by quantum mechanical and
entropic effects.  Such stability and the subsequent collapse has been
observed in clouds of degenerate $^7$Li \cite{REF:randy}.  The collapse of
the Bose gas shares many features of the gravitational collapse of cold
interstellar hydrogen, described by the Jeans instability
\cite{REF:jeans}; related instabilities occur in supercooled vapors. 
Theoretical studies of the attractive Bose gas, typically numerical, have
been limited to zero \cite{REF:zeroT} or very low \cite{REF:finiteT}
temperature.  Here we give a simple analytic description of the region of
stability and the threshold for collapse, valid from zero temperature to
well above the Bose condensation transition, and thus provide a consistent
global picture of the instability.  As we show, the non-condensed
particles play a signficant role in the collapse at finite temperature
\cite{noncond}.  Indeed in the normal state, collapse occurs even when no
condensate exists. 

\begin{figure}[tb]
  \epsfxsize=\columnwidth\nobreak
  \centerline{\epsfbox{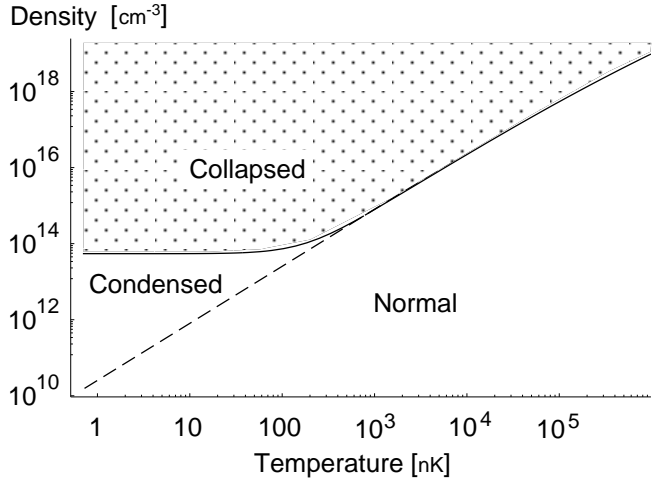}}\nobreak
 \vskip2mm\nobreak
 \caption{
Phase diagram, density $n$ [cm$^{-3}$] versus temperature $T$ [nK], for a
small cloud of attractive bosons with spatial extent $L=3.15 \mu$m, scattering
length $a_s = -1.45$nm, and atomic weight 7 (see [1]).  Note the logarithmic
scales.  To apply these results to harmonically trapped gases, the density $n$
should be interpreted as the central density in the trap.  The solid line
separates the unstable (shaded) region from the stable region.}
\label{FIG:phase}
\end{figure}

    Our results are summarized in the phase diagram in Fig.~\ref{FIG:phase},
which shows three regions:  normal, Bose condensed, and collapsed.  This third
region is not readily accessible experimentally; the system becomes unstable
at the boundary (the solid line in the figure).  The attractive interactions,
which drive the instability, have a characteristic energy $V=\hbar^2 (a_s
n)/m$ per particle, where $\hbar$ is Planck's constant, $a_s$ ($=-1.45$nm for
$^7$Li) is the s-wave scattering length, $n$ is the density, and $m$ is the
particle mass.  At low temperature the only stabilizing force is quantum
pressure, arising from the zero-point motion of the particles, with a
characteristic energy $E_Q=\hbar^2/mL^2$, where $L$ is the size of the cloud.
At higher temperatures, thermal pressure, resulting from the decrease in
entropy with decreasing volume of a system, also helps to stabilize the cloud.
The characteristic energy scale for thermal pressure is $E_T=k_B T$, where $T$
is the temperature.  Naively, the line of collapse is where $V$ becomes larger
than both $E_T$ and $E_Q$.  Quantum pressure rapidly decreases as the system
size increases, and in the thermodynamic limit the entire condensed phase is
swallowed up by the collapsed phase (cf.~Fig.~2).  Due to the finite system
size the line separating the normal and condensed phase is not a phase
transition, but rather is a rapid cross-over.

    The zero temperature stability of a finite boson cloud can be understood
in terms of the Bogoliubov excitation spectrum of a uniform gas
\cite{REF:bog}:
\begin{equation}\label{EQ:bog}
    \omega^2 = \left( \frac {k^2} {2m} \right)^2 +  g n \frac {k^2}{m},
\end{equation}
where $g=4 \pi\hbar^2a_s/m$.  In the attractive case, $g<0$, all long
wavelength modes with $k^2/2m < 2|g|n$ have imaginary frequencies and are
unstable.  A system of finite size $L$ has modes only for $k>2\pi/L$ and will
have an excitation spectrum similar to Eq.~(\ref{EQ:bog}) for short
wavelengths.  Thus one expects that a zero temperature attractive Bose gas
will be stable if $|g|n<\hbar^2\pi^2/mL^2$ \cite{REF:dim}.  We extend this
analysis to finite temperature by calculating the zero frequency
density-density response function $\chi(k)$.  The onset of an instability in
the longest wavelength mode is signaled by a divergence in $\chi(k=2\pi/L)$.

    The main role of finite temperature, $T$, is to supplement quantum
pressure with thermal pressure.  In the limit $T\gg T_c$, quantum effects are
negligible, and the line of collapse becomes the spinodal line of a classical
liquid-gas phase transition, as characterized by Mermin \cite{REF:mermin}.  To
illustrate the physics of this non-degenerate regime we look for an
instability in the uniform gas at zero wavevector, $k=0$, calculating the
compressibility in the Hartree approximation \cite{REF:hartree}, and
effectively neglecting finite size effects such as quantum pressure.  Within
this approximation the density is given by the self-consistent solution of
\begin{equation}\label{EQ:density}
  n = \int\frac{d^3p}{(2\pi\hbar)^3}
    \frac{1}{e^{\beta (\varepsilon(p)-\mu)}-1},
\end{equation}
with Hartree quasiparticle energies $\varepsilon(p)=p^2/2m + g n$.  Here
$\mu$ is the chemical potential, and $\beta = 1/k_BT$.  In the classical
limit, this equation becomes $n= e^{\beta(\mu- g n)}/\Lambda_T^3$, where the
thermal wavelength is $\Lambda_T= (2\pi\hbar^2/m k_B T)^{1/2}$.  The
compressibility is proportional to the zero wavevector density-density
response function, $\chi(k=0)=-\partial n/\partial \mu$.  We introduce the
``bare'' response, $\chi_0(k=0)= -\left(\partial n/
\partial\mu\right)_\varepsilon$, where the $\varepsilon$ are held fixed; in
the classical limit $\chi_0(0)=-\beta n$.  The full response $\chi(0)$, as in
the random phase approximation (RPA), is
\begin{equation}\label{EQ:hf}
 \chi(k) = \frac{\chi_0(k)}{1- g \chi_0(k)};
\end{equation}
in the classical limit $\chi(0)=-\beta n/(1+g\beta n)$.  The collapse of
the attractive system ($g<0$) is signaled by a divergence of the
compressibility.  In the classical limit, this divergence occurs when $|g|
n=k_BT$ \cite{REF:dim}.  The approximations made here break down when
$|\mu|\sim\hbar^2/mL^2$ and if applied outside their range of validity would
erroneously predict that the instability towards collapse prevents Bose
condensation from occurring.

    In the experiments on $^7$Li, the atoms are held in a magnetic trap with a
harmonic confining potential $V(r)=\frac{1}{2} m \omega^2 r^2$, with
$\omega\approx 2\pi\times 145 s^{-1}$, and have a highly inhomogeneous density
profile \cite{REF:freq}.  In connecting the theory of the uniform gas with the
experiment, $n$ should be interpreted as the density at the center of the
trap, and the system size $L$ as effectively $\sqrt{\hbar/m\omega}$
($=3.2\mu$m).  As long as $k_B T$ is larger than the trap energy
$\hbar\omega\approx 7 nK$, the geometry of the trap should play no significant
role in the collapse.

    We describe the system more generally using the response function
$\chi(k)$ derived from the shielded potential approximation \cite{KB}, as used
by Sz\'epfalusy and Kondor to study the critical behavior of collective
modes \cite{REF:szef}.  This approach, a simple example of the dielectric
formalism, generates an excitation spectrum which is conserving
\cite{REF:cons} and gapless \cite{REF:griffin}.  At zero temperature it yields
the Bogoliubov spectrum shown in Eq.~(\ref{EQ:bog}), and above $T_c$ it
becomes the standard RPA.  Analysis in terms of Bogoliubov quasiparticles is
left to a future paper.

    The susceptibility has the general form, Eq.~(\ref{EQ:hf}), with the
appropriate polarization part $\chi_0$.  Within the shielded potential
approximation $\chi_0$ is the response of the non-interacting system.  This
response can be broken up into two parts, $\chi_0^s$ and $\chi_0^r$,
corresponding to the condensate and non-condensate contributions
\cite{REF:sr}, which at finite frequency are given by
\begin{mathletters}
\begin{eqnarray}
\chi_0^r(k,\omega) &=& \int\frac{d^3q}{(2\pi)^3}
 \frac{ f(q-k/2)-f(q+k/2)}
      {\omega-(\varepsilon_{q+k/2} -\varepsilon_{q-k/2})},\\
\chi_0^s(k,\omega) &=&
  \frac{n_0}{\omega-\varepsilon_k}-\frac{n_0}{\omega+\varepsilon_k}.
\end{eqnarray}
\end{mathletters}
Here $n_0$ is the condensate density, $\varepsilon_k=k^2/2m$ is the free
particle spectrum, and the Bose factors $f(k)$ are given by
$(e^{-\beta(\varepsilon_k-\mu)}-1)^{-1}$.

\begin{figure}[tb]
  \epsfxsize=\columnwidth\nobreak
  \centerline{\epsfbox{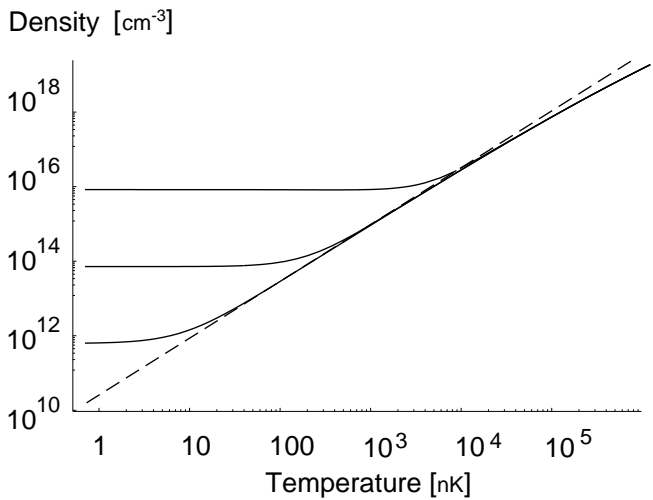}}\nobreak
 \vskip2mm\nobreak
\caption{
    Scaling of the instability threshold with system size.  Solid lines show
the maximum stable density $n_{\rm max}$ for a given temperature.  From the
top, the system size is $L$ = 0.3, 3, 30 $\mu$m.  The other parameters are the
same as Fig.~\ref{FIG:phase}.  The dashed line indicates the Bose-Einstein
condensation transition.  Note the three scaling regimes; at low temperature,
$n_{\rm max}\sim L^2$, near the critical temperature $n_{\rm max}\sim
L^{3/2}$, and at high temperatures $n_{\rm max}$ is independent of $L$ (see
Eqs.~(5) and (6)).
}\label{FIG:scaling}
\end{figure}

    Expanding $\chi_0^r$ in the small parameter $k\Lambda_T$ we derive for
$T>T_c$,
\begin{mathletters}
\begin{eqnarray}\label{EQ:aboveT}
g\chi_0^r(k) &=&  -2\frac{a_s}{\Lambda_T}
    \left[\frac{4\pi}{k\Lambda_T}
          \arctan\left(\frac{1}{2}\left|\frac{\epsilon_k}{\mu}\right|^{1/2}
               \right)\right.
             \\\nonumber&& \quad\left.
+ g_{1/2}(e^{\beta\mu})
-\left|\frac{\pi}{\beta\mu}\right|^{1/2}
+{\cal O}(k\Lambda_T)
\right],\\
g\chi_0^s(k) &=&0,\\
n&=& \frac{1}{\lambda_T^3} g_{3/2}(e^{\beta\mu}).
\end{eqnarray}
\end{mathletters}
where $g_\nu(z) \equiv \sum_j z^j/j^\nu$ is the polylogarithm function.
For chemical potential $\mu$ much larger in magnitude than $k_B T$, the system
is in the classical regime, and Eq.~(\ref{EQ:aboveT}) reduces to
$g\chi_0^r=-\beta g n$, as was found in the Hartree approach.  Below $T_c$ the
chemical potential of the non-interacting system vanishes and the response
functions are simpler:
\begin{mathletters}
\begin{eqnarray}\label{EQ:belowT}
g\chi_0^r(k) &=& -\frac{4\pi^2 a_s}{k \Lambda_T^2}
+ {\cal O}((k\Lambda_T)^{0}),\\
g\chi_0^s(k) &=& - 16\pi\frac{a_s n_0}{k^2},\\
n&=& n_0 + \frac{1}{\Lambda_T^3} \zeta(3/2),
\end{eqnarray}
\end{mathletters}
with $\zeta(\nu)=g_\nu(1)$, the zeta function.  Using these expressions we
calculate the spinodal line separating the stable and unstable regions of
Fig.~\ref{FIG:phase} by setting $k=2\pi/L$ and solving the equation
\begin{equation}
 \chi^{-1}\propto 1-g(\chi_0^s+\chi_0^r)=0.
\end{equation}
Note that the condensate response $\chi_0^s$ scales as $L^2$, whereas below
$T_c$ the noncondensate response $\chi_0^r$ scales as $L$.  For realistic
parameters, $L$ is the largest length in the problem, so the condensate
dominates the instability except when $n_0$ is much smaller than $n$.  In
Fig.~\ref{FIG:scaling} we plot the line of instability for various $L$.

\begin{figure}[tbh]
  \epsfxsize=\columnwidth\nobreak
  \centerline{\epsfbox{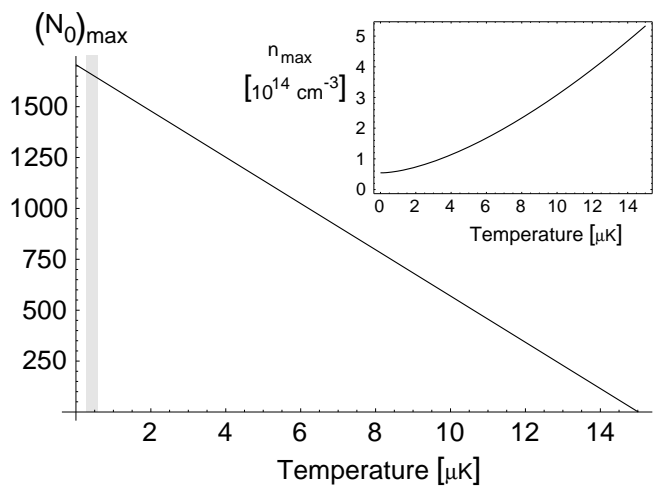}}\nobreak
 \vskip2mm\nobreak
    \caption{Maximum number of condensed particles as a function of
temperature for the same parameters as in Fig.~\ref{FIG:phase}.  Present
experiments are performed in the shaded region.  The inset shows the maximum
total density for the same temperature range.
}\label{FIG:cond}
\end{figure}

    From the above equations we calculate the maximum stable value of the
condensate density $n_0$:
\begin{equation}
(n_0)_{\rm max} = \frac{\pi}{4L^3} \left(\frac{L}{|a|}
      - \frac{L^2 m k_B T}{\hbar^2} \right);
\end{equation}
$(n_0)_{\rm max}$ decreases linearly with temperature, eventually
vanishing at $T = \hbar^2/mL|a|$.
Conversely, the maximum total density
increases rather quickly with temperature.
Using parameters from
the Rice experiments \cite{REF:randy}, we find
\begin{equation}\label{EQ:rnmax}
(N_0)_{\rm max} = L^3 (n_0)_{\rm max} = 1706-T/(8.8 nK).
\end{equation}
This result, plotted in Fig.~\ref{FIG:cond}, is consistent with the
experiments, and agrees qualitatively with numerical mean-field calculations
\cite{REF:finiteT}.  Quantitative agreement requires a more sophisticated
treatment of the geometry.  To verify the structure in Eq.~(\ref{EQ:rnmax})
and to map out the phase diagram in Fig.~\ref{FIG:phase}, future condensate
experiments will need to be performed at temperatures of several $\mu$K, an
order of magnitude hotter than the current 100 to 300 nK experiments.  At
these temperatures the density needed for collapse is so high that inelastic
two and three body processes seriously restrict the lifetime of the condensate
\cite{REF:lifetime}.  Thus it may become necessary to work with softer traps
(larger $L$), where the line of collapse lies at lower densities (see
Fig.~\ref{FIG:scaling}).

    The formalism used here to discuss the collapse of a gas with attractive
interactions also describes domain formation in spinor condensates, and
gives a nice qualitative understanding of experiments at MIT \cite{REF:spin}
in which optically trapped $^{23}$Na is placed in a superposition of two spin
states.  Although all interactions in this system are repulsive, the two
different spin states repel each other more strongly than they repel
themselves, resulting in an effective attractive interaction.  The collapse
discussed earlier becomes, in this case, an instability towards phase
separation and domain formation.  The equilibrium domain structure is
described in \cite{REF:spintheory}.  Here we focus on the formation of
metastable domains.

    The ground state of sodium has hyperfine spin $F=1$.  In the experiments
the system is prepared so that only the states $|1\rangle=|F=1,m_F=1\rangle$
and $|0\rangle=|F=1,m_F=0\rangle$ are involved in the dynamics.  The effective
Hamiltonian is then
\begin{equation}\label{EQ:ham}
H = \int\!\!d^3r\, \frac{\nabla \psi_i^\dagger \cdot
                         \nabla \psi_i}{2m}
+ V(r)\psi_i^\dagger\psi_i + \frac{g_{ij}}{2}
 \psi^\dagger_i\psi^\dagger_j\psi_j\psi_i.
\end{equation}
where $\psi_i$ ($i=0,1$) is the particle destruction operator for the
state $|i\rangle$; summation over repeated indices is assumed.  The effective
interactions, $g_{ij}=4\pi\hbar^2 a_{ij}/m$, are related to scattering
amplitudes $a_{F=0}$ and $a_{F=2}$, corresponding to scattering in the singlet
($F_1+F_2=0$) and quintuplet ($F_1+F_2=2$) channel by
\cite{REF:spin,REF:spintheory,REF:jason}:
\begin{mathletters}
\begin{eqnarray}
 \tilde a \equiv a_{11}=a_{01}=a_{10}  = a_{F=2}, \\
 \delta a \equiv   a_{11}-a_{00} = (a_{F=2}-a_{F=0})/3.
\end{eqnarray}
\end{mathletters}
Numerically, $\tilde a = 2.75$nm and $\delta a = 0.19$nm.  We introduce
$\tilde g=4\pi\hbar^2\tilde a/m$ and $\delta g=4\pi\hbar^2\delta a/m$.  In the
mean field approximation, the interaction in Eq.~(\ref{EQ:ham}) becomes a
function of $n_{m=0}$ and n, the density of particles in the
state $|0\rangle$ and the total density, respectively:
\begin{equation}
\langle H_{int}\rangle =
\int\!\!d^3r\,\left(\frac{\tilde g}{2} n^2 - \frac{\delta g}{2}
n_{m=0}^2\right),
\end{equation}
which shows explicitly the effective attractive interaction.  Initially
the condensate is static withdensity $n=10^{14} {\rm cm}^{-3}$, and all
particles in state $|1\rangle$.  A radio frequency pulse places half of the
atoms in the $|2\rangle$ state without changing the density profile.
Subsequently the two states phase separate and form domains from 10 to 50
$\mu$m thick.  The trap plays no role here, so we can neglect $V(r)$ in
Eq.~(\ref{EQ:ham}) and consider a uniform cloud.

    Linearizing the equations of motion with an equal density of particles in
each state, we find two branches of excitations corresponding to density and
spin waves \cite{REF:gold},
\begin{mathletters}
\begin{eqnarray}
\omega_{ph}^2 &=& \left(\frac{k^2}{2 m}\right)^2
+ \tilde{g} n \frac{k^2}{m}+
{\cal O}(\delta g),\\\label{EQ:spin}
\omega_{sp}^2 &=&
\left(\frac{k^2}{2 m}\right)^2
- \delta g\, n \frac{k^2}{4 m}
+ {\cal O}((\delta g)^2).
\end{eqnarray}
\end{mathletters}
Since $\delta g>0$ we see the appearance of spin excitations with
imaginary frequency.  The mode with the largest imaginary frequency grows most
rapidly, and the width of the domains formed should be comparable to the
wavelength $\lambda$ of this mode.  By minimizing Eq.~(\ref{EQ:spin}) one
finds $\lambda= \sqrt{2\pi/n\,\delta a} =10 \mu$m, in rough agreement with the
observed domain size.

    The authors are grateful to the Ecole Normale Sup\'erieure in Paris, and
the Aspen Center for Physics, where this work was carried out.  We owe special
thanks to Eugene Zaremba and Dan Sheehy for critical comments.  The research
was supported in part by the Canadian Natural Sciences and Engineering
Research Council, and National Science Foundation Grant No.  PHY98-00978.
This work was facilitated by the Cooperative Agreement between the University
of Illinois at Urbana-Champaign and the Centre National de la Recherche
Scientifique.


\begin{references}
\bibitem[*]{REF:EJM}
Electronic address: {\tt emueller@physics.uiuc.edu\/}
\bibitem[\dagger]{REF:GB}
Electronic address: {\tt gbaym@uiuc.edu\/}
\bibitem{REF:randy}
C.A. Sackett, H.T.C. Stoof, and
      R.G. Hulet, Phys. Rev. Lett. {\bf80} 2031, (1998);
C.A. Sackett, C.C. Bradley, M. Welling, and R.G. Hulet, Appl.
      Phys. B {\bf65} 433, (1997);
C.C. Bradley,
      C.A. Sackett, and R.G. Hulet, Phys. Rev. Lett. {\bf 78} 985, (1997);
C.C. Bradley, C.A. Sackett, and
      R.G. Hulet, Phys. Rev. A {\bf 55} 3951, (1997).

    \bibitem{REF:jeans} For elementary discussions of the Jeans instability
see, e.g., R. Bowers, and T. Deeming {\em Astrophysics II; Interstellar Matter
and Galaxies} (Jones and Bartlett, Sudbury, MA, 1984).

    \bibitem{REF:zeroT} P.A.  Ruprecht, M.J. Holland, K. Burnett, and M.
Edwards, Phys.Rev.  A {\bf 51}, 4704 (1995); G. Baym and C. J. Pethick, 
Phys. Rev.  Lett.  {\bf 76} 6 (1996).

    \bibitem{REF:finiteT} M.J. Davis, D.A.W.  Hutchinson, and E. Zaremba,
J. Phys. B {\bf 32}, 3993 (1999);  T. Bergeman, Phys.  Rev.  A {\bf 55},
3658 (1997); M.  Houbiers and H.T.C.  Stoof, Phys.  Rev.  A {\bf 54}, 5055
(1996); P.A.  Ruprecht, M.J. Holland, K. Burnett, and M. Edwards, Phys. 
Rev.  A {\bf 51}, 4704 (1995). 

    \bibitem{noncond} Previous theoretical studies \cite{REF:finiteT}
focused on the stability of the condensate at very low temperatures
and in a regime where the density of non-condensed particles is much
smaller than the condensate density.  These studies found that the
non-condensed particles play a very small role in the collapse.  For
example, Stoof et al. concluded that the non-condensed particles are
involved in the collapse only in so far as they change the geometry of the
effective potential felt by the condensate.  In this low temperature
regime our results are consistent with these previous works.

    \bibitem{REF:bog} N.N.  Bogoliubov, J. Phys.  USSR, {\bf 11}, 23 (1947)

    \bibitem{REF:dim} Up to a factor of order unity, this result can be
derived through dimensional analysis.

    \bibitem{REF:mermin} N.D.  Mermin, Ann.  Phys.  {\bf 18}, 421, 454 (1962);
{\bf 21}, 99 (1963).

    \bibitem{REF:hartree} We neglect exchange here to aid comparison with the
RPA, which extends Hartree theory to finite wavevector and frequency.  It is
straightforward to add exchange to the formalism, but such details needlessly
obscure the discussion here.

    \bibitem{REF:freq} The trap used is slightly asymmetric, and the frequency
quoted here is the geometric mean of the three frequencies;
see~\cite{REF:randy}.

    \bibitem{KB} L.P.~Kadanoff and G. Baym, {\it Quantum Statistical
Mechanics} (W.A.~Benjamin, New York, 1962).

    \bibitem{REF:szef} P. Sz\'epfalusy and I. Kondor, Ann.  Phys.  {\bf 82}, 1
(1974).

    \bibitem{REF:cons} G. Baym, Phys.  Rev.  {\bf 127}, l39l (1962).

    \bibitem{REF:griffin} A. Griffin, {\em Excitations in a Bose-Condensed
Liquid} (Cambridge University Press, NY, 1993).

    \bibitem{REF:sr} As used by Sz\'epfalusy and Kondor \cite{REF:szef}, $s$
and $r$ denote singular and regular.

    \bibitem{REF:lifetime}  At a temperature of $5\mu K$, a $3.15\mu$m
cloud of $^7$Li collapses at a density of
$n=1.39 \times 10^{14}{\rm cm}^{-3}$.
At such a high density the dominant decay mechanism is 3-body
collisions, giving a lifetime $\tau = (G_3\, n^2)^{-1}$.  Using
the theoretical estimate \cite{REF:decay},
$G_3=2.6\times10^{-28}$, we find that $\tau=200 ms$.
At $T=10\mu K$ the lifetime is only $\tau=40 ms$.

    \bibitem{REF:decay} A. J. Moerdijk, H. M. J. M. Boesten,
and B. J. Verhaar, Phys. Rev. A {\bf 53}, 916 (1996);
C. A. Sackett, J. M. Gerton, M. Welling, and R. G. Hulet,
Phys. Rev. Lett. {\bf 82}, 876 (1999).

    \bibitem{REF:spin} H.-J.  Miesner, D. M. Stamper-Kurn, J. Stenger, S.
Inouye, A. P. Chikkatur, and W. Ketterle, Phys.  Rev.  Lett.  {\bf 82}, 2228
(1999).

    \bibitem{REF:spintheory} T. Isoshima, K. Machida, and T. Ohmi,
preprint,{\tt cond-mat/9905182}.


    \bibitem{REF:jason} T.-L.~Ho, Phys.  Rev.  Lett.  {\bf 81}, 742 (1998).

    \bibitem{REF:gold} Similar excitation spectra are found in E.V.~Goldstein
and P. Meystre, Phys.  Rev.  A {\bf 55}, 2935 (1997).

\end{references}
\end{document}